**Article title: Assessing biophysical and socio-economic impacts of climate change on avian biodiversity.**

**Authors:** Simon Kapitza[1*], Pham Van Ha[2], Tom Kompas[2,3], Nick Golding[1], Natasha C. R. Cadenhead[1,4], Payal Bal[1] and Brendan A. Wintle[1,4]

**Author affiliations:**

[1]Quantitative and Applied Ecology Group, School of BioSciences, The University of Melbourne, Parkville, Victoria, 3010, Australia

[2]Crawford School of Public Policy, Lennox Crossing, Australian National University, Australian Capital Territory, 2601, Australia

[3]Centre of Excellence for Biosecurity Risk Analysis, School of BioSciences, The University of Melbourne, Parkville, Victoria, 3010, Australia

[4]NESP Threatened Species Recovery Hub, University of Melbourne and University of Queensland, St Lucia, Queensland, 4072, Australia

**Corresponding author:** Simon Kapitza, Quantitative and Applied Ecology Group, School of BioSciences, The University of Melbourne, Victoria, 3010, Australia, simon.kapitza@unimelb.edu.au

# Assessing biophysical and socio-economic impacts of climate change on avian biodiversity

Simon Kapitza, Pham Van Ha, Tom Kompas , Nick Golding, Natasha C. R. Cadenhead, Payal Bal and Brendan A. Wintle

## Abstract

Climate change threatens biodiversity directly by influencing biophysical variables that drive species' geographic distributions and indirectly through socio-economic changes that influence land use patterns, driven by global consumption, production and climate. To date, no detailed analyses have been produced that assess the relative importance of, or interaction between, these direct and indirect climate change impacts on biodiversity at large scales. Here, we apply a new integrated modelling framework to quantify the relative influence of biophysical and socio-economically mediated impacts on avian species in Vietnam and Australia. We find that socio-economically mediated impacts on suitable ranges are largely outweighed by biophysical impacts, but global shifts of production are likely to result in adverse impacts on habitats worldwide. By translating economic futures and shocks into spatially explicit predictions of biodiversity change, we now have the power to analyse in a consistent way outcomes for nature and people of any change to policy, regulation, trading conditions or consumption trend at any scale from sub-national to global.

## Significance statement

We present a novel framework for integrated macro-economic, land use, and biodiversity change modelling that permits quantitative analysis of questions critical to land use and biodiversity outcomes under broader socio-economic narratives, but also very specific policy scenarios. We are now in a position to analyse the impacts of diverse domestic and international policy settings on land use and biodiversity, including changes to trade agreements and other economic shocks. Applying this new framework, we provide a first assessment of the relative magnitude of socio-economically and biophysically mediated climate change impacts on biodiversity in Vietnam and Australia.

**Introduction**

Climate change affects biodiversity through a multitude of pathways. There is pervasive evidence that climate change directly affects environmental conditions that are related to the climatic niches of many taxa, with the potential of significant shifts in their distributional ranges or even the total extinction of species(1, 2). However, climate change also affects biodiversity through indirect human-mediated impacts: it drives the loss of livelihoods and displacement(3) and affects food and commodity production systems through its impacts on land productivity and human health(4, 5) and environmental suitability for different land uses(6, 7). Resulting global transitions of land use patterns are set to drive habitat conversion and may have dramatic impacts on biodiversity(8–10). While there are some examples of studies examining synergistic effects of land use and climate change on species (11, 12), large-scale assessments of biodiversity change in response to climate change have tended to look only at direct impacts of climate change on biophysical conditions or habitat loss and fragmentation alone(8). Analyses that couple direct biophysical impacts on species with indirect socio-economic impacts via consumption, commodity, and land use change are sorely needed to fill important gaps in our knowledge of interactions between land use and climate change(10), to foster a more holistic understanding of the impacts of climate change, and to support the design of cross-sectoral adaptation and mitigation strategies(13).

No single model of drivers of change in biodiversity and ecosystem services can capture all relevant dynamics at a high level of detail and there is an increasing awareness of the urgency to consider interactions between direct and indirect drivers of change under future scenarios to characterise prospects and management options for biodiversity and ecosystem services(13). Coupling demographic, economic and biophysical models has the potential to advance understanding and improve representation of synergies between direct and indirect drivers in biodiversity modelling, and to discover non-linear system behaviours that may not be apparent when considering drivers in isolation(13).

Here, we contribute to the recent advances in integrated assessment modelling(14–17) by applying an integrated modelling framework to compare the relative influence of direct biophysical and indirect socio-economic climate change impacts on the distribution and extent of suitable ranges for avian species in Vietnam and Australia (Fig. 1).

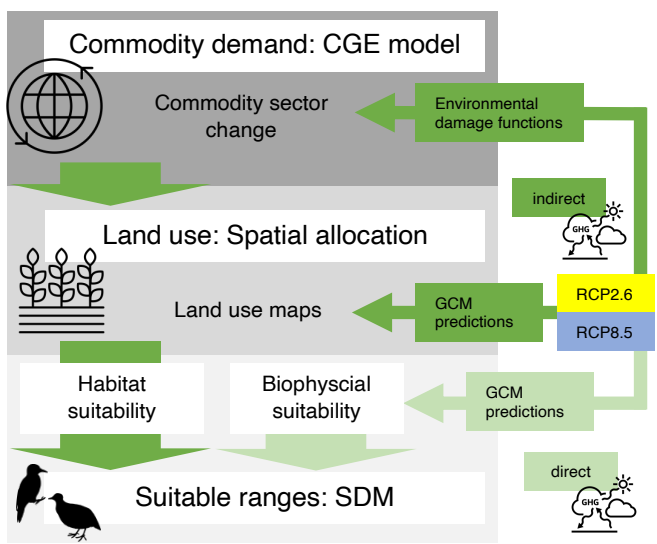

**Fig. 1 | Overview of the modelling framework to capture interactions between direct and indirect drivers of biodiversity change under climate change scenarios.** We included two Representative Concentration pathways RCP2.6

and RCP8.5 to characterize the plausible extremes of climate change. Dark green arrows represent the indirect pathway of climate change impacts on suitable ranges. Light green arrows indicate the direct pathway of climate change impacts on ecological suitability. Icons from thenounproject.com.

Recent advances in computable general equilibrium (CGE) modelling(18, 19) bring unprecedented power to parametrise the impacts of climate change on commodity consumption and production patterns at very high commodity and temporal resolution across the global economy. We combine this economic modelling power with state-of-the-art land use change modelling to spatially downscale commodity demand changes caused by climate change(4) into changes of land use patterns. The spatial realisation of changing land use patterns varies with changes in the suitability of land for particular uses and is thereby also driven by climate change(7, 20). Commodity demand changes are projected annually and land use predicted in 10-year time steps, producing decadal time-series maps of land use. Maps are integrated with climate change predictions into a biodiversity impact assessment using species distribution models (SDMs)(21–24) . SDMs, fitted to current climate, land use, and other environmental variables (Supplementary Table 1) are extrapolated to conditions in 2070 under a range of climate and land use scenarios. Predictions of relative likelihood of occurrence are thresholded to examine changes in the ecologically suitable ranges for 1282 bird species in Vietnam and Australia(21–24).

# Methods

**Study area.** We focussed our analysis on Vietnam and Australia because the countries provide unique socio-economic contexts, while hosting a similar number of bird species that are vulnerable, endangered or critically endangered(25, 26). Due to the country's small size and limited number of occurrences, SDMs for Vietnam were built using data from a 30 x 30-degree tile that comprises large parts of Southeast Asia. This enabled us to capture the occurrence of bird species present in Vietnam, across a much broader range of environmental variables, enabling better prediction of likely occurrence under future climates.

**Climate Change.** We chose two alternative representative concentration pathways (RCP) that represent two extremes of the expected radiative forcing levels, RCP2.6 and RCP8.5(27). For each pathway, we acquired the 2070 predictions of the first 19 bioclimatic variables downscaled as part of WorldClim Version 1.4(28) from 15 GCM of Coupled Model Intercomparison Phase 5 (CMIP5)(29). In order to account for variation of GCM predictions under different models, we determined the cell-wise first and third quartiles, as well as the medians of each of the 19 variables across the 15 GCM (Supplementary Table 2 for a complete list of the GCM used in this analysis).

Main results were derived by predicting land use and species distributions under the medians of these variables. We also predicted both land use and species distributions under the first and third quartiles to approximate the range of outcomes for species across all 15 GCM (Supplementary Fig. 1). In CGE models, we included the parametrization of both climate change pathways proposed by Roson & Satori(4).

**CGE models.** We developed an inter-temporal Global Trade and Analysis Project (GTAP) model(30) to simulate changes in production under different climate change scenarios. CGE models use input-output-tables derived from national economic census data. These tables represent the inputs required in each economic sector to produce outputs and meet household and government demands (both nationally and internationally), which in turn is affected by prices and thus supply. Sectors are linked within each national economy, but also between economies. Producers in each country can produce various commodities to sell domestically (to household and government) and internationally to foreign households and governments. The households and governments generate their income from selling (to producers) productive input factors (land, capital, labour, etc.) and through taxes. In our version of GTAP, the total land area (land endowment) from which allocations are made to crop sectors (land requirements) can be changed in the baseline. Therefore, land supply is not necessarily fixed, as is the case in most other GTAP models. Estimations within GTAP are carried out relative to this baseline supply and we convert relative changes in agricultural land requirements to absolute changes in cropland by using their respective shares in the total harvested area for a by-sector-weighting of the average relative change of all classes. This weighted average change is applied on the current area under cropland to derive a future value. This means there is a direct proportional link between changes in land requirements and changes in the total area of agricultural land and the total area under agricultural land can change at the expense or to the benefit of other classes.

Our inter-temporal GTAP model uses the GTAP 9 database(31), which is subdivided into 139 regions and 57 commodity sectors(31) and extends the GTAP model by replacing the recursive dynamic module of the current GTAP model with a forward-looking dynamic (or inter-temporal) module, where the producer can optimise profits overtime(32, 33). More than just a trade model, the inter-temporal GTAP model allows optimal investment behaviours, in which producers in each country can adjust their decisions based on the impacts from both past and foreseeable future events. Agents in the model can react to future threats long before their full realisation(33). This makes the model a perfect tool for the simulation of future phenomena like climate change.

Climate change impacts are modelled in our GTAP model following the work by Roson & Satori(4), in which impacts are realised as shocks to land supply and agricultural and labour productivity. The reduction in endowments of productive land and productivity negatively affect the production of commodities. Agricultural commodities are expected to be the most affected. With production shrinking more in some commodities than others, the price will adjust to balance the demand and supply of commodities. As a result, we will see a substitution effect between domestically produced products and their competitive imports along with a substitution effect in factors of production (such as land), balancing demands between sectors.

Unlike the Kompas et al.(33) approach, which relied on a one-step simulation approach, here we apply a multi-step simulation approach allowing the shocks to be applied into smaller successive intervals combined with extrapolation techniques to further enhance the simulation accuracy (see Horridge et al. (34) and Pearson(35) for the details on multi-steps CGE solution methods). The solution of the inter-temporal GTAP model in this paper has been carried out within a parallel computing platform(19, 36) with the use of PETSC (37–39) and HSL (40) libraries.

**Land use models.** We reclassified a global land-use map to 8 land use classes (urban, cropland, herbaceous ground vegetation, shrubland, open canopy forest, closed canopy forest and wetlands and barren land) (Supplementary Table 1 for full list of data sources). To aggregate projected changes in land requirements for agricultural sectors to a single agricultural land use class (cropland), we fixed the area contribution of each agricultural sector at 2016 levels(41) (Supplementary Table 3) and calculated a weighted average change in land requirements for all cropland classes in each time step. Changes in urban land were estimated using estimates of urban population changes(42) and adjusting the amount of land under this class, assuming that urban population density remains steady through time. Future applications of this work will establish links between land-use classes related to forestry and livestock-raising, as has been demonstrated recently(17).

We predicted land use maps under both pathways in 10-year time steps, using an R implementation (R package 'lulcc'(43)) of the Conversion of Land Use and its Effects at Small regional extents (CLUE-S) model proposed by Verburg et al.(44). First, we determined the local suitability for different land uses through logistic regression of land use against the linear combination of a range of biophysical and socio-economic drivers in Generalised Linear Models (GLMs), from 15,000 randomly sampled pixels in each region (Supplementary Table 1 for a detailed list of data, Supplementary Figure 3 for effect sizes of predictors in each GLM). The selection of variables for land use suitability models was based on work by Verburg et al(45). ). Correlation analysis eliminated highly correlated predictor pairs (Spearmen's rank correlation coefficient ≥ 0.7), always keeping the predictor whose highest correlation with any other remaining predictor was smaller, to maximise independent information retained in the final set. The final predictor sets were checked against literature (46, 47). We discarded a small number of predictors using cross-validated Lasso penalisation in the 'glmnet' R-package(48) and used the reduced predictor sets to build GLM and predict to future timesteps by interpolating GCM-predicted WorldClim variables in 10-year time steps under RCPs 2.6 and 8.5 (Supplementary Table 2 for used GCM models). GLM predictions produced probability maps that represent the landscape's potential suitability for each land use class, under consideration of changes in the included variables. Transitions between classes were restricted according to a transition matrix that specified which transitions were possible (Supplementary Table 4). We specified conversion elasticities of each class (the amount by which land-use can shift without changing the total area it occupies) based on literature(43, 44).

To satisfy the CGE-projected changes in land endowments, changes were allocated in areas with the highest suitability for each land use, until estimated land area demands were met(49). Competition between land uses is handled in CLUE-S by allocating the land-use with the highest predicted suitability in a given cell, accounting for conversion elasticity and allowed transitions. We masked category I and II protected areas (7), precluding these areas from land-use changes. Since there was no CGE-modelled future demand for herbaceous ground vegetation and shrubland, as well as the forest classes, the overall amount of area allocated to those land uses was simply what was left over from the uses prescribed by agricultural and other demands. The proportional allocation between each of these residual categories was determined based on their mean predicted suitability in the landscape. All land use simulations were made using GCM-predicted first and third quartiles, as well as the medians of bioclimatic variables.

**Species distribution models.** Correlative species distribution models (SDM) can predict responses of species to changing environmental conditions by extrapolating from the covariate space in which they were observed(21–24). The MaxEnt software package(50) (ver. 3.3.3k) was used to fit SDMs for 656 bird species in Australia and 739 bird species in Vietnam, using presence-only data from the Global Biodiversity Information Facility (GBIF)(51). We filtered records to retain observations from 1950-2018 and records with more than or equal to 20 occurrence points across the model-fitting area. We included a range of biophysical, topographic and socio-economic predictors as well as land use (Supplementary Table 1). Correlation analysis eliminated highly correlated predictor pairs (Spearmen's rank correlation coefficient ≥ 0.7), using the same method as for land use predictors. We ensured through literature review that the final predictor sets were ecologically meaningful to avian species across taxa (52–55) at our aspired scale. We kept 9 predictors for Australia and 10 predictors for Vietnam, including 5 and 6 climate predictors respectively.

Sampling bias is a pervasive issue particularly affecting presence-only data that is often sampled opportunistically. We accounted for sampling bias by estimating the intensity of sampling effort in response to demographic and topographic variables(56), and using this map of sampling effort to probabilistically select background points. By selecting background points proportional to sampling bias the effect of sampling effort on the location of presence records is largely eliminated as a form of bias(57). Variables used in the bias models were protected area status, distance to roads, distance to built-up areas and roughness.

Predictions were made using the estimated quartiles and medians of bioclimatic variables and the according land use maps that were also predicted under quartiles and medians. We controlled overfitting by determining the permutation importance of predictors and dropping predictors with a value < 1%. Test AUC were estimated via 5-fold cross-validation of each model and final models built on all available records. Species for which only uninformative models were fitted (AUC < 0.7) were excluded (58).

We recorded the log ratio of the respective number of cells with relative likelihoods predicted above MaxEnt's MaxSSS threshold(59) (where the sum of model sensitivity and specificity is maximised) between present (2018) and future time step (2070) as a measure of change . In Australia, we constrained this change estimation for each species to bioregions containing records of the species, and adjacent bioregions(60).

**Software and data.**

All data preparation and modelling for land use and SDMs was conducted in R(61), using packages 'lulcc'(43) for land use simulations and 'dismo'(62) for MaxEnt models. All analyses and spatial predictions of the land use model and SDM were performed at 0.5 arc-minute resolution; approximately 1 km at the equator. SDM building and predictions were computationally expensive and required up to 50 GB of working memory on 12 parallel cores.

# Results

## Direct biophysical impacts dominate changing range sizes.

For birds in both regions, we forecast major declines in ecologically suitable ranges, with severity of loss scaling with the severity of climate change (Fig. 2). Under RCP 8.5, a much higher number of species would be expected to experience decreases of more than half of their present ecologically suitable range compared with RCP 2.6, although variation in responses is also greater, indicated by the much wider spread of points (Fig. 2a, b). In Australia, mean suitable range decline under both pathways is not predicted to be as severe as in Vietnam and a smaller number of species is predicted to lose more than half of their suitable range. For both Vietnamese and Australian birds, predicting only under the indirect (land use change) effects of climate change results in little change to mean predicted outcomes for species (Fig 2a,b), though some threatened species are predicted to lose significant suitable range within their current range due to indirect climate change impacts (see below). Mean predictions under combined direct and indirect effects do not differ to any notable degree from those made under direct biophysical effects only. Predictions under the first and third quartiles of bioclimatic variables across 15 Global Circulation Models (GCMs) show the same trends identified in the main results (Supplementary Fig. 1).

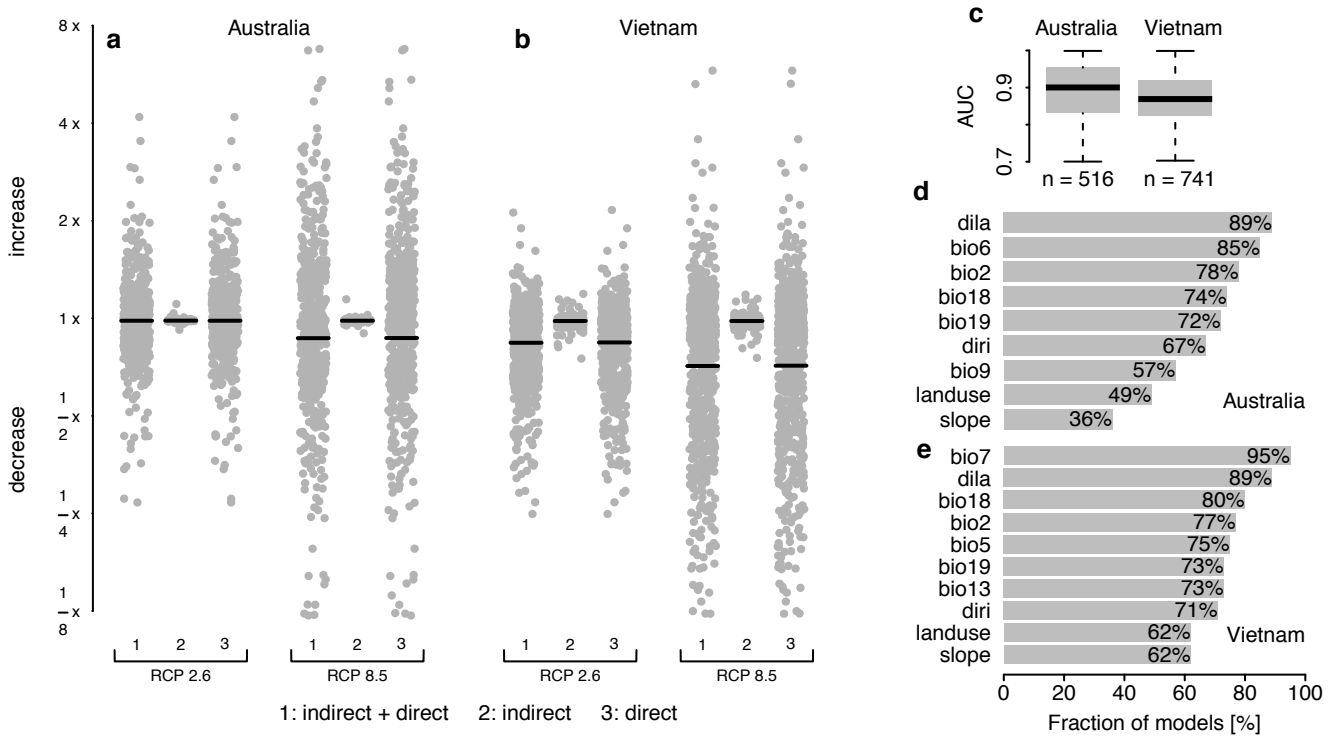

**Fig. 2 | Predicted changes in species' ecologically suitable ranges. a, b -** Illustration of multiplicative changes in species' ecologically suitable ranges between present (2018) and 2070 for Australia and Vietnam respectively, under three treatments (1) "indirect + direct" (combined biophysical and socio-economic impacts of climate change), (2) "indirect" (net socio-economic impacts) and (3) "direct" (net biophysical impacts). Each point corresponds to a species, black bars are means of ecologically suitable range changes across all species. **c -** A summary of cross-validated test Area Under the Receiver Operating Characteristics Curve values (AUCs)(63) of models in the two regions as well as the respective number of models (*n*) retained (AUC > 0.7)(58). AUC provides a measure of a model's discriminatory performance in terms of how well test predictions discriminate between occupied and unoccupied locations(58, 63). **d, e -** Fractions of models in which a predictor was used. Full names and definitions of all predictors can be found in Supplementary Table 1.

SDMs for 1436 bird species were used in the analysis of the direct and indirect impacts of climate change on biodiversity. Discriminatory performance of the SDMs was assessed using cross-validated AUCs which varied between 0.7 and 1.0 with a mean of 0.90 in Vietnam and 0.87 in Australia (Fig 2c), indicating very high discriminatory performance. We discarded models for 179 species with AUC < 0.7 (see Methods). The predictor variables retained in the highest fraction of models were distance to lakes (*dist lakes*) in Australia and annual temperature range (*bio7*) in Vietnam. These are followed by *dist lakes* and precipitation of warmest quarter (*bio18*) in Vietnam, and by minimum temperature of the coldest week (*bio6*) and mean diurnal temperature range (*bio2*) in Australia. In Australia, *land use* was retained in about half the models. The very minor indirect (via *land use*) impact predictions arise because the changes in commodity demand predicted by the CGE model did not result in significant changes to land use in both regions (see below).

**Land use changes in response to climate change vary by region.**

The total output of most agricultural crop sectors in both regions was predicted to decrease more with increasing climate change. In particular, in Vietnam, sectors such as oil seeds and plant-based fibres shrink by up to 20% under RCP 8.5 (Fig. 3a). The land requirements for each sector generally increase in proportion to the overall output of each sector. This is due to climate change impacts on crop yields as parametrised in the CGE-model: reductions in land productivity mean that more land is required to maintain sector outputs. Accordingly, in both countries, even while total outputs tend to decrease, land requirements of agricultural sectors remain approximately the same, or increase slightly (Fig. 3a).

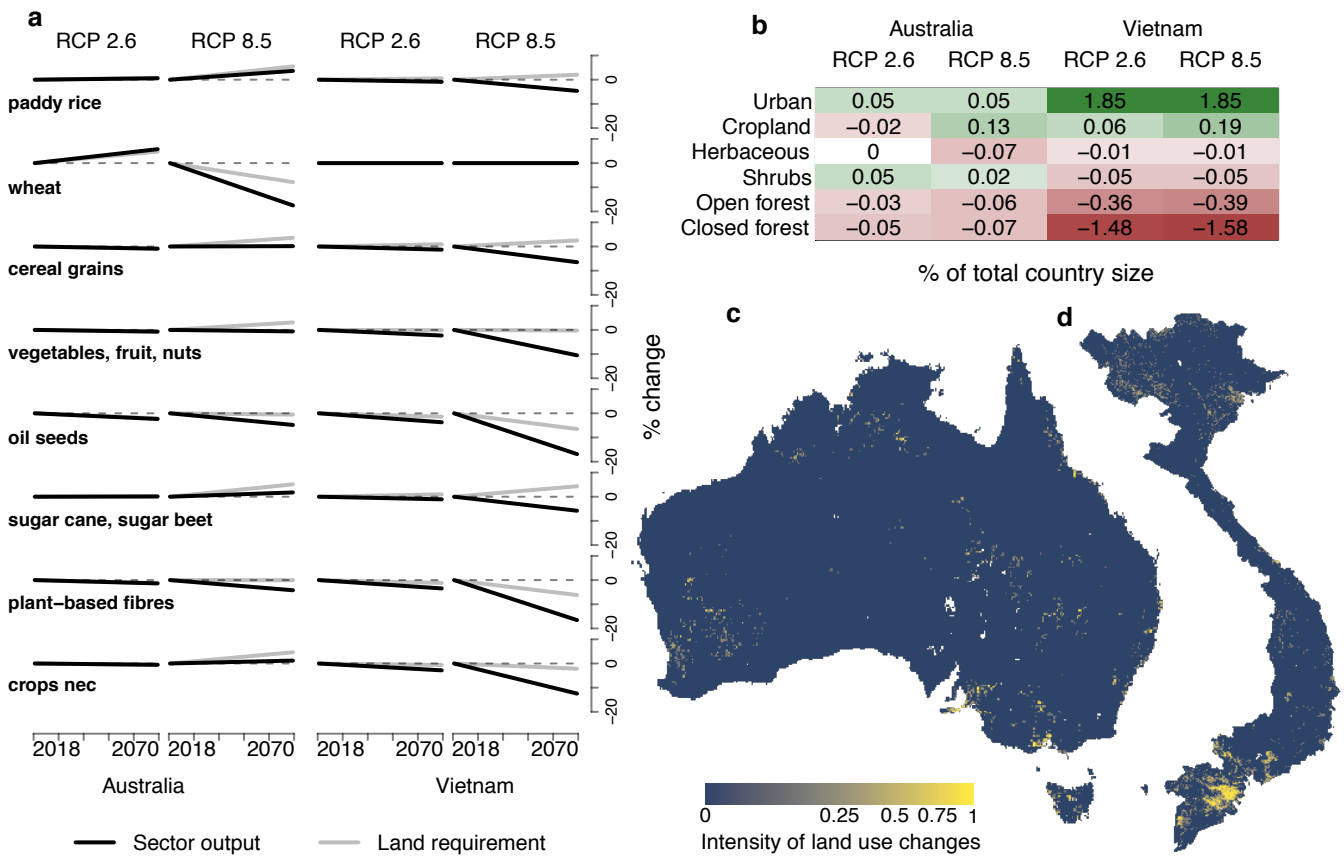

**Fig. 3 | CGE and land use model results. a,** Future projections of commodity sector output and sector land endowments (the area required to produce output of a sector) from CGE model under RCP 2.6 and RCP 8.5. **b,** Illustration of the percentage change of each land use in response to GTAP projections of crop sectors in **a** and FAO urban population projections, grouped by

country and RCP, relative to the whole country size. **c-d**, Intensity of predicted land use changes under indirect effects of RCP 8.5 in **c** Australia and **d** Vietnam. These maps are derived by aggregating predicted land use changes between any two classes under the indirect impacts of RCP 8.5 by factor 3.

The changes in land requirements for crop lead to an increase in cropland of <0.5% of the total land area in both regions under RCP 8.5 and a very slight decrease in Australia under RCP 2.6 (Fig. 3b). Increases in urban land in both countries were modelled on FAOSTAT estimates of urban population growth(41). In Australia, land use changes occur locally and are concentrated in coastal areas along the north-east, south and west of the continent, although some changes also occur further inland (Fig. 3c). In Vietnam, land use change is higher overall, with a particular concentration of change in the central-southern and northern coastal areas of the country, that also approximately coincide with the country's major river deltas (Fig. 3d). Given that the distributions of most species are constrained, aggregated, and not random, small percentage changes in land use at the national scale still have significant impacts on some species locally (Fig. 4a, c). For example, species losing more than 10% of their currently suitable range under indirect impacts of RCP 8.5 in Vietnam include the vulnerable chestnut-necklaced partridge (*Arborophila charltonii*) and the near-threatened yellow-billed nuthatch (*Sitta solangiae*). These declines are highly localised and predominantly occur in the centre-south of the country (Fig. 4c). Direct climate change impacts are more severe: 324 and 362 species lose at least 10% of their suitable ranges under direct impacts of RCP 2.6 and RCP 8.5 respectively, with areas particularly affected across taxa under RCP 8.5 in the northern highlands and the central eastern parts of the country (Fig. 4d). Among the species losing more than 95% of their current suitable range under the direct impacts of RCP 8.5 are the Chinese thrush (*Turdus mupinensis)* and the critically endangered white-rumped vulture (*Gyps bengalensis*) (Fig. 4cd.

In Australia, no species was found to lose more than 10% of its currently suitable range under indirect climate change impacts, although the black-throated whipbird (*Psophodes nigrogularis*) loses more than 5%. A higher number of species are affected by the direct impacts of climate change, with areas predicted to suffer particularly high suitable range declines along the southern and eastern coasts, the southwest and the southeast of the continent. In Australia, 188 and 230 species are expected to lose more than 10% of their suitable range under RCP 2.5 and RCP 8.5 respectively.

Amongst the Australian species losing more than 95% of their suitable range under the direct impacts of RCP 8.5 are the kalkadoon grasswren (*Amytornis ballarae*) and the Australasian pitpit (*Anthus Australis*) and a number of other species now categorised as of least concern (Fig. 4b). This highlights the potential dangers of climate change to species that we do not yet consider under threat, but for which extinction debts are accruing(64).

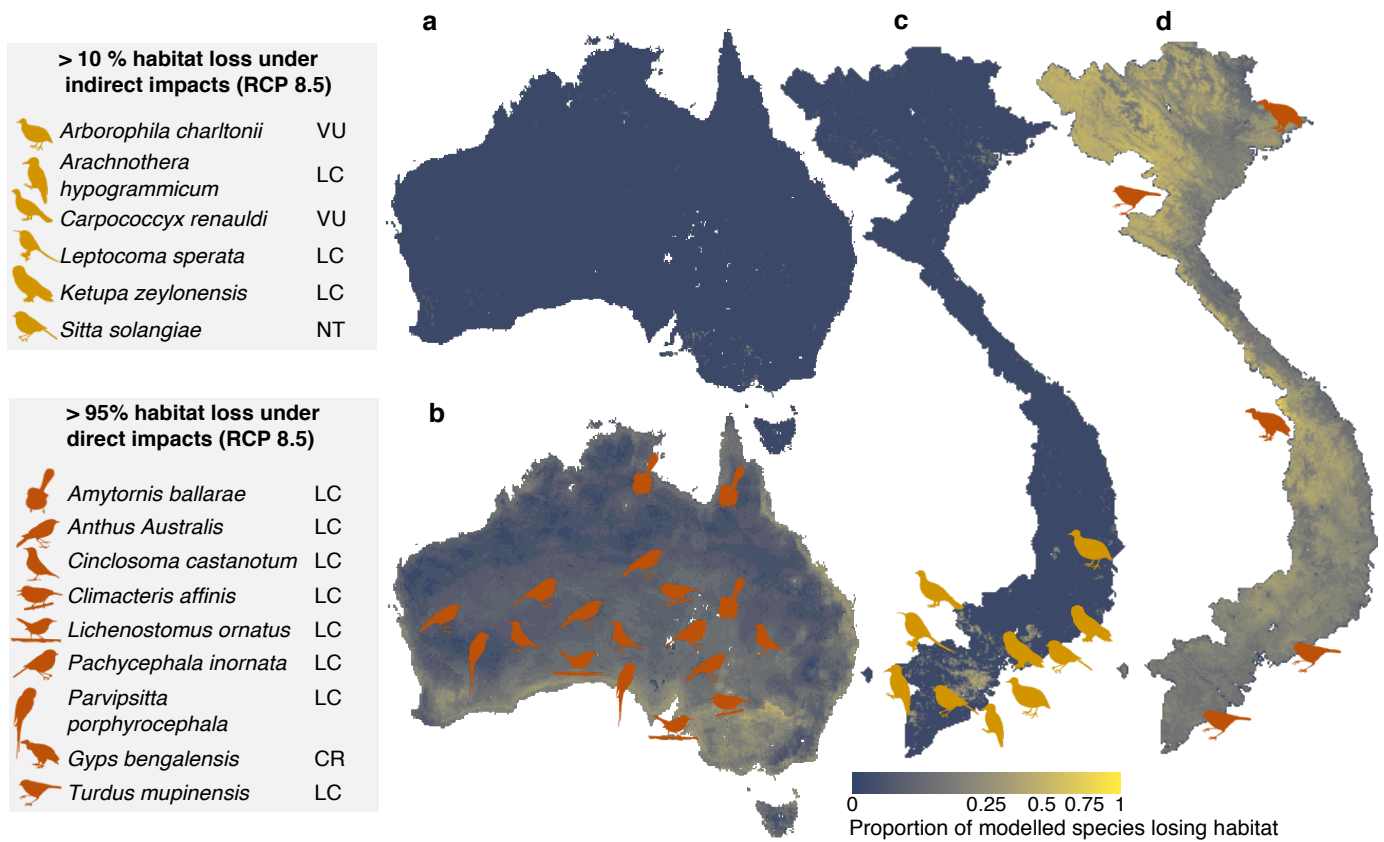

**Fig. 4 | Mapping of habitat loss under RCP 8.5. a–d**, The proportion of avian species predicted to lose ecologically suitable range across Australia (**a, b**) and Vietnam (**c, d**) under the indirect (**a, c**) and direct (**b, d**) climate change impacts under RCP 8.5. Cell shading indicates the proportion of species predicted to lose suitable range in each cell. This identifies areas of declines in species' suitable ranges from either indirect or direct impacts. The icons indicate locations of suitable range declines for severely affected species that lose more than 10% (indirect) and more than 95% (direct) of their suitable ranges overall. IUCN conservation status is given alongside taxonomic names (LC – least concern; VU – vulnerable; NT – near threatened; EN – endangered; CR – critically endangered)(65). Icon credit - http://phylopic.org.

The expected direct impacts of climate change impacts on many taxa are well researched and documented (i.e. increased extinction risks across taxa with accelerated climate change(66), northward shifts of bird distributions in Great Britain under climate change(53) and responses of bird abundance to climate change in the United States and Europe(67)). Our findings largely agree with these trends. In both Australia and Vietnam, climate change is likely to have extensive detrimental impacts on the climatically suitable ranges of birds. For many species, suitable ranges decline with increasing severity of climate change (Fig. 2) and under RCP 8.5, 24% of species analysed (in Vietnam) show likely declines in suitable ranges of greater than 50%, increasing their extinction risk in the country severely. Our analysis shows that subject to the assumptions of this work, the relative contribution of direct, biophysical impacts of climate change on biophysical suitability in our study area outweighs the contribution of indirect socio-economic impacts on habitat suitability via global commodity markets and resulting land use change, also taking into account the fact that climate change impacts on the suitability of land for particular uses. In Vietnam and Australia, bird species appear to be more severely impacted by the direct influence of changing climates than by its indirect impacts via commodity demand and land use.

**Alternative economic and land use futures may drive more profound biodiversity impacts.** Better understanding of climate change impacts on commodity demand and supply, and how those changes impact biodiversity should remain a research priority. Our results are valid for avian taxa in Australia and Vietnam under a number of assumptions about how commodity demand and supply, land use and biodiversity interact to deliver outcomes predicted by our integrated model. In our assessment framework, we follow a top-down modelling approach; within the architecture of our CGE model, climate change affects global demand and supply of many land-based commodities, requiring sector outputs as well as requirements of land to each sector to increase or decrease. However, mapped land use changes corresponding to changes in land endowments to different commodity sectors do not feed back into the CGE model. The inclusion of such feedbacks would increase the realism of both CGE and land use predictions, but detailed knowledge of local production systems and commodity markets are required to accurately parameterise such a model, and such models are computationally expensive(68).

We predict economic change under two climate change scenarios, keeping all other aspects of the global economy at current baseline values. This way, we could capture and isolate the effect of climate change on the economy. However, this approach omits other socio-economic processes that could affect supply and demand, such as population growth, changes to economic growth, energy efficiency, and shifts in social demands. These other factors may impact habitat and biodiversity through agricultural expansion, deforestation or urbanisation. While this study was designed to assess the net effects of direct and indirect climate change impacts on species as a first case study introducing our integrated assessment framework, these factors will be incorporated in future iterations that include an even more comprehensive CGE parametrization (i.e. full CGE baseline scenarios with socio-economic pathway narratives(69) and integration of climate models in CGE analysis) and through improvements to current CGE methods by including, for example, stochastic effects of natural disasters in the CGE modelling.

**Biodiversity model assumptions may underplay potential indirect impacts.** We chose not to produce SDMs for species with less than 20 occurrence records, to avoid the inflation of AUC for range-restricted species and species with very low prevalence (70) and to assure sufficient discrimination between presences and background points (71, 72). Rare or spatially restricted species can be more vulnerable to localised impacts such as habitat loss(73), but these effects are difficult to capture when biodiversity data are poor. We assumed unlimited dispersal ability in Vietnam and dispersal ability constrained to bioregions adjacent to those containing observation records in Australia. Disconnected patches of potential habitat outside of observed ranges (but within adjacent bioregions in Australia) were counted in future predictions, regardless of whether those areas were functionally linked (by suitable or traversable habitats) to the observed range and thus within the dispersal range of species. This may lead to an over-estimation of habitat utilisation and a commensurate underestimation of both direct and indirect impacts of climate change, particularly for taxa with a low dispersal ability that rely on small pockets of habitat within their range and are unable to reach disconnected patches of potential habitat. The importance of connectivity as a key component of habitat structure is well known(74) and crucial for population viability in many species with low dispersal ability(75, 76). The parametrization of species' dispersal ability and explicit modelling of landscape structure in response to land use change would allow for the inclusion of these fragmentation effects. This may be particularly important when our framework is extended to non-avian species .

**Discussion**

**Exported biodiversity impacts**. While we found that total agricultural sector outputs decrease in both Vietnam and Australia, decreases in land productivity mean that land use in production for some agricultural commodities were predicted to increase slightly. We assumed a global economic equilibrium in which commodities can be substituted through trade between regions, thus implying that global demand for land-based commodities is serviced by regions that benefit from a comparative advantage under climate change. Where comparative advantage is due to increases in land productivity (land use efficiency), additional land may not be required to increase outputs. However, where this advantage is due to other economic mechanisms and not driven by the cost of converting additional land for production, more land may be allocated to agricultural or other commodity production, increasing habitat loss. For example, in Canada, our CGE model predicted an increase of wheat sector output by over 37% under RCP 8.5, while land endowments increase by only 14% due to increases in land use efficiency. In India, wheat output is estimated to increase by 8% under RCP 8.5, while land endowments to the wheat sector increase by 6 %, suggesting much lower land use efficiency in India than in Canada (see Supplementary Fig. 2 for a global, country-wise mapping of projected changes in sector outputs and land endowments of the wheat sector). Despite lower land use efficiency, wheat production in India still grows, because growth is economically feasible as long as it is not limited by factors arising from the sector's context in domestic and international markets. In both countries, increases in land use lead to agricultural expansion, but in Canada more wheat can be produced per unit land and areas lost to wheat farming are likely to be much smaller per produced unit than in India. Nonetheless, if wheat production occurs in parts of Canada that were previously in, for example, natural prairie, then significant biodiversity losses may occur. Our framework provides in-depth insight into the links between sectors and regions and allows for a better understanding of global shifts in land requirements, enabling the fine-scale identification of hotspots for production, agricultural expansion and ultimately habitat destruction under consideration of the global economic processes.

**High potential for global analyses of economic impacts on biodiversity.** In this first implementation of our framework we could capture and quantify principal relationships between climate change, the global economy, land use and avian habitat. Future uses of our approach could include regional and global biodiversity assessments following individual policy shocks, such as the introduction or abolishment of taxes or international trade deals, or could seek to capitalize on existing narratives of socio-economic futures and climate change pathways (so-called *Shared Socio-economic Pathways*)(69) to parametrise climate adaptation policies, sustainable development goals and other aspects of socio-political transitions within the CGE modelling. Expanding consideration of biodiversity to include non-avian taxa and explicitly dealing with the role of connectivity and dispersal will enable a more comprehensive assessment of biodiversity impacts under socio-economic change. A key feature of our approach is that it provides opportunity to downscale country-level commodity demands to spatial explicit land use changes and biodiversity impacts, enabling a more meaningful analysis of the habitat and biodiversity implications of economic shocks or the implications of trade than have previously been possible.

Better integration of models and scenarios of biodiversity is required to guide evidence-based climate adaptation strategies and to chart progress toward sustainable development goals(77). Our approach to integrating economic, land use and biodiversity values into a single model capable of high resolution, spatially-explicit predictions of land use and biodiversity outcomes provides information in a form that can be used directly by planners and managers. While spatial predictions of biodiversity and land use change have been available for decades, being able to place these predictions coherently in a global economic context is a new and exciting development that will bring a new level of relevance and realism to predictions in the eyes of policy and decision makers.

## Acknowledgements

This work received funding under the Australian Research Council Discovery grant DP170104795. The authors acknowledge contributions made throughout the research phase by B. Bryan, P. Lentini, H. Kujala and J. Elith.

## Author contributions

SK led the modelling with contributions from HPV, TK, NG, NC, BW. BW, TK and SK conceptualised the analysis framework. SK led the writing of the manuscript with contributions from all authors.

## Competing interests

The authors declare no competing interests.

## Data Availability

Sources for data used in land use and species distribution modelling are listed in Supplementary Information. Direct download links to these data sets are available in the code repository accompanying this study (see below). We provide outputs of the CGE, land-use and species distribution models in a data repository (DOI: 10.5281/zenodo.3517724). The GTAP database that underpins GGE modelling is available from GTAP under license. However, we provide input files and commented CGE modelling code that contains details of parameters settings for global economic models and detailed commodity demand output tables for each of the scenarios modelled upon publication. Components of the GTAP database can be provided on request.

## Code Availability

All R-code for land use and species distribution modelling is available online (DOI: 10.5281/zenodo.2703618). The tablo-code for the CGE modelling will be provided in the code repository accompanying this study upon publication.

## References

1. IPBES, Summary for policymakers of the global assessment report on biodiversity and ecosystem services of the Intergovernmental Science-Policy Platform on Biodiversity and Ecosystem Services. Advance unedited version. Ngo, H. T.; Guèze, M.; Agard, J.; Arneth, A.; Balvanera, P.; Brauman, K.; Butchart, S.; Chan, K.; Garibaldi, L.; Ichii, K.; Liu, J.; Subramanian, S. M.; Midgley, G.; Miloslavich, P.; Molnár, Z.; Obura, D.; Pfaff, A.; Polasky, S.; Purvis, A.; Razzaque, J.; Reyers, B.; Chowdhury, R. R.; Shin, Y-J; Visseren- Hamakers, I.; Willis, K.; Zayas, C.. Secretariat of the Intergovernmental

Science-Policy Platform on Biodiversity and Ecosystem Services, Bonn, Germany. (2019).

2. M. J. Struebig, et al., Targeted Conservation to Safeguard a Biodiversity Hotspot from Climate and Land-Cover Change. Current Biology **25**, 372–378 (2015).

3. A. J. McMichael, R. E. Woodruff, S. Hales, Climate change and human health: present and future risks. The Lancet **367**, 859–869 (2006).

4. R. Roson, M. Sartori, Estimation of Climate Change Damage Functions for 140 Regions in the GTAP 9 Database. Journal of Global Economic Analysis **1**, 78–115 (2016).

5. R. S. J. Tol, "Who Benefits and Who Loses from Climate Change?" in Handbook of Climate Change Mitigation and Adaptation, W.-Y. Chen, T. Suzuki, M. Lackner, Eds. (Springer New York, 2014), pp. 1–12.

6. A. Veldkamp, L. O. Fresco, CLUE: a conceptual model to study the Conversion of Land Use and its Effects. Ecological Modelling **85**, 253–270 (1996).

7. P. H. Verburg, K. P. Overmars, Combining top-down and bottom-up dynamics in land use modeling: Exploring the future of abandoned farmlands in Europe with the Dyna-CLUE model. Landscape Ecology **24**, 1167–1181 (2009).

8. C. S. Mantyka-pringle, et al., Climate change modifies risk of global biodiversity loss due to land-cover change. Biological Conservation **187**, 103–111 (2015).

9. T. H. Oliver, M. D. Morecroft, Interactions between climate change and land use change on biodiversity: Attribution problems, risks, and opportunities. Wiley Interdisciplinary Reviews: Climate Change **5**, 317–335 (2014).

10. T. Newbold, et al., Climate and land-use change homogenise terrestrial biodiversity, with consequences for ecosystem functioning and human well-being. Emerging Topics in Life Sciences, ETLS20180135 (2019).

11. J. F. Brodie, Synergistic effects of climate change and agricultural land use on mammals. Front Ecol Environ **14**, 20–26 (2016).

12. M. Brambilla, P. Pedrini, A. Rolando, D. E. Chamberlain, Climate change will increase the potential conflict between skiing and high-elevation bird species in the Alps. J. Biogeogr. **43**, 2299–2309 (2016).

13. IPBES, Summary for policymakers of the methodological assessment of scenarios and models of biodiversity and ecosystem services of the Intergovernmental Science-Policy Platform on Biodiversity and Ecosystem Services. S. Ferrier, K. N.; Ninan, P.; Leadley, R.; Alkemade, L. A.; Acosta, H. R.; Akçakaya, L.; Brotons, W. W. L.; Cheung, V. Christensen, K. A.; Harhash, J.; Kabubo-Mariara, C.; Lundquist, M.; Obersteiner, H. M.; Pereira, G.; Peterson, R.; Pichs-Madruga, N.; Ravindranath, C.; Rondinini and B. A. Wintle (eds.). Secretariat of the Intergovernmental Science-Policy Platform on Biodiversity and Ecosystem Services, Bonn, Germany. (2016).

14. D. Leclere, et al., "Towards pathways bending the curve of terrestrial biodiversity trends within the 21st century" (International Institute Of Applied System Analysis, 2018) (October 16, 2018).

15. T. Newbold, Future effects of climate and land-use change on terrestrial vertebrate community diversity under different scenarios. Proceedings of the Royal Society B: Biological Sciences **285**, 20180792 (2018).

16. R. P. Powers, W. Jetz, Global habitat loss and extinction risk of terrestrial vertebrates under future land-use-change scenarios. Nature Climate Change (2019) https:/doi.org/10.1038/s41558-019-0406-z (March 12, 2019).

17. A. Marques, et al., Increasing impacts of land use on biodiversity and carbon sequestration driven by population and economic growth. Nature Ecology & Evolution (2019) https:/doi.org/10.1038/s41559-019-0824-3 (March 6, 2019).

18. P. V. Ha, T. Kompas, H. Thi, M. Nguyen, C. Hoang, Building a Better Trade Model to Determine Local Effects : A Regional and Intertemporal GTAP Model (2016).

19. P. Van Ha, T. Kompas, Solving intertemporal CGE models in parallel using a singly bordered block diagonal ordering technique. Economic Modelling **52**, 3–12 (2016).

20. R. Fuchs, M. Herold, P. H. Verburg, J. G. P. W. Clevers, A high-resolution and harmonized model approach for reconstructing and analysing historic land changes in Europe. Biogeosciences **10**, 1543–1559 (2013).

21. C. R. Lawson, J. A. Hodgson, R. J. Wilson, S. A. Richards, Prevalence, thresholds and the performance of presence-absence models. Methods in Ecology and Evolution **5**, 54–64 (2014).

22. B. A. Wintle, J. Elith, J. M. Potts, Fauna habitat modelling and mapping: A review and case study in the Lower Hunter Central Coast region of NSW. Austral Ecology **30**, 719–738 (2005).

23. B. a. Wintle, et al., Ecological–economic optimization of biodiversity conservation under climate change. Nature Climate Change **1**, 355–359 (2011).

24. C. D. Thomas, Climate change and extinction risk. Nature **430**, 25 (2004).

25. BirdLife International, Country profile: Vietnam (2018) (April 12, 2018).

26. BirdLife International, Country Profile: Australia (2018) (April 12, 2018).

27. D. P. van Vuuren, et al., The representative concentration pathways: An overview. Climatic Change **109**, 5–31 (2011).

28. R. Hijmans, S. Cameron, J. Parra, P. Jones, A. Jarvis, WORLDCLIM - a set of global climate layers (climate grids), version 1.4.

29. K. E. Taylor, R. J. Stouffer, G. A. Meehl, An Overview of CMIP5 and the Experiment Design. Bulletin of the American Meteorological Society **93**, 485–498 (2012).

30. T. Hertel, Global Trade Analysis: Modeling and applications (Center for Global Trade Analysis, Department of Agricultural Economics, Purdue University, 1997).

31. A. Aguiar, B. Narayanan, R. McDougall, An Overview of the GTAP 9 Data Base. Journal of Global Economic Analysis **1**, 181–208 (2016).

32. P. Van Ha, T. Kompas, H. T. M. Nguyen, C. H. Long, Building a better trade model to determine local effects: A regional and intertemporal GTAP model. Economic Modelling **67**, 102–113 (2017).

33. T. Kompas, V. H. Pham, T. N. Che, The Effects of Climate Change on GDP by Country and the Global Economic Gains From Complying With the Paris Climate Accord. Earth's Future (2018) https:/doi.org/10.1029/2018EF000922 (August 28, 2018).

34. J. M. Horridge, M. Jerie, D. Mustakinov, F. Schiffmann, GEMPACK manual, GEMPACK Software, ISBN 978-1-921654-34-3 (2018).

35. K. R. Pearson, Solving nonlinear economic models accurately via a linear representation, Working paper No. IP-55. Victoria University, Centre of Policy Studies (1991).

36. T. Kompas, P. V. Ha, The 'curse of dimensionality' resolved: The effects of climate change and trade barriers in large dimensional modelling. Economic Modelling, 24 (2018).

37. S. Balay, et al., "PETSc users manual, Technical Report ANL-95/11 - Revision 3.11" (Argonne National Laboratory, 2019).

38. S. Balay, et al., PETSc Web page (2019).

39. S. Balay, W. D. Gropp, L. C. McInnes, B. F. Smith, "Efficient management of parallelism in object oriented numerical software libraries" in Modern Software Tools in Scientific Computing, E. Arge, A. M. Bruaset, H. P. Langtangen, Eds. (Birkhaeuser Press, 1997), pp. 163–202.

40. HSL, A collection of fortran codes for large scale scientific computation. The HSL Mathematical Software Library (2013).

41. Food and Agriculture Organization of the United Nations (FAO), FAOSTAT Statistics Database (2017) (September 18, 2018).

42. World Bank Group, Population Estimates and Projections. http://data.worldbank.org/data-catalog/population-projection-tables (2016).

43. S. Moulds, W. Buytaert, A. Mijic, An open and extensible framework for spatially explicit land use change modelling: The lulcc R package. Geoscientific Model Development **8**, 3215–3229 (2015).

44. P. H. Verburg, et al., Modeling the spatial dynamics of regional land use: The CLUE-S model. Environmental Management **30**, 391–405 (2002).

45. P. H. Verburg, T. A Veldkamp, J. Bouma, Land use change under conditions of high population pressure: The case of Java. Global Environmental Change **9**, 303–312 (1999).

46. P. H. Verburg, C. J. E. Schulp, N. Witte, A. Veldkamp, Downscaling of land use change scenarios to assess the dynamics of European landscapes. Agriculture, Ecosystems and Environment **114**, 39–56 (2006).

47. P. H. Verburg, G. H. J. De Koning, K. Kok, A. Veldkamp, J. Bouma, A spatial explicit allocation procedure for modelling the pattern of land use change based upon actual land use. Ecological Modelling **116**, 45–61 (1999).

48. J. Friedman, T. Hastie, R. Tibshirani, Regularization Paths for Generalized Linear Models via Coordinate Descent. J. Stat. Soft. **33** (2010).

49. R. Fuchs, M. Herold, P. H. Verburg, J. G. P. W. Clevers, A high-resolution and harmonized model approach for reconstructing and analysing historic land changes in Europe. Biogeosciences **10**, 1543–1559 (2013).

50. S. J. Phillips, R. P. Anderson, R. E. Schapire, Maximum entropy modeling of species geographic distributions. Ecological Modeling **190**, 231–259 (2006).

51. GBIF, GBIF data portal. http://www.gbif.net/ (2016).

52. S. J. Goetz, M. Sun, S. Zolkos, A. Hansen, R. Dubayah, The relative importance of climate and vegetation properties on patterns of North American breeding bird species richness. Environmental Research Letters **9**, 034013–034013 (2014).

53. S. Gillings, D. E. Balmer, R. J. Fuller, Directionality of recent bird distribution shifts and climate change in Great Britain. Global Change Biology **21**, 2155–2168 (2015).

54. R. Maggini, et al., Assessing species vulnerability to climate and land use change: The case of the Swiss breeding birds. Diversity and Distributions **20** (2014).

55. C. L. Coxen, J. K. Frey, S. A. Carleton, D. P. Collins, Species distribution models for a migratory bird based on citizen science and satellite tracking data. Global Ecology and Conservation **11**, 298–311 (2017).

56. J. Stolar, S. E. Nielsen, Accounting for spatially biased sampling effort in presence-only species distribution modelling. Diversity and Distributions **21**, 595–608 (2015).

57. S. J. Phillips, et al., Sample selection bias and presence-only distribution models: Implications for background and pseudo-absence data. Ecological Applications **19**, 181–197 (2009).

58. R. Baldwin, Use of Maximum Entropy Modeling in Wildlife Research. Entropy **11**, 854–866 (2009).

59. C. Liu, G. Newell, M. White, On the selection of thresholds for predicting species occurrence with presence-only data. Ecology and Evolution **6**, 337–348 (2016).

60. A. Morán-Ordóñez, J. J. Lahoz-Monfort, J. Elith, B. A. Wintle, Evaluating 318 continental-scale species distribution models over a 60-year prediction horizon: what factors influence the reliability of predictions? Global Ecology and Biogeography, 1–14 (2016).

61. R Development Core Team, R: A language and environment for statistical computing (Foundation for Statistical Computing, 2008).

62. R. J. Hijmans, S. Phillips, J. Leathwick, J. Elith, Package "dismo" (2011) (August 20, 2018).

63. A. Jiménez-Valverde, Insights into the area under the receiver operating characteristic curve (AUC) as a discrimination measure in species distribution modelling. Global Ecology and Biogeography **21**, 498–507 (2012).

64. M. Kuussaari, et al., Extinction debt: a challenge for biodiversity conservation. Trends in Ecology & Evolution **24**, 564–571 (2009).

65. IUCN, The IUCN Red List of Threatened Species. Version 2018-2 (2018).

66. M. C. Urban, Accelerating extinction risk from climate change. Science **348**, 571–573 (2015).

67. P. A. Stephens, et al., Consistent response of bird populations to climate change on two continents. Science **352**, 84–87 (2016).

68. B. A. Bryan, et al., Land-use and sustainability under intersecting global change and domestic policy scenarios: Trajectories for Australia to 2050. Global Environmental Change **38**, 130–152 (2016).

69. D. P. van Vuuren, T. R. Carter, Climate and socio-economic scenarios for climate change research and assessment: Reconciling the new with the ol. Climatic Change **122**, 415–429 (2014).

70. A. S. J. van Proosdij, M. S. M. Sosef, J. J. Wieringa, N. Raes, Minimum required number of specimen records to develop accurate species distribution models. Ecography **39**, 542–552 (2016).

71. P. A. Hernandez, C. Graham, L. L. Master, D. L. Albert, The effect of sample size and species characteristics on performance of different species distribution modeling methods. Ecography **29**, 773–785 (2006).

72. M. S. Wisz, et al., Effects of sample size on the performance of species distribution models. Diversity and Distributions **14**, 763–773 (2008).

73. A. Guisan, et al., Using Niche-Based Models to Improve the Sampling of Rare Species. Conservation Biology **20**, 501–511 (2006).

74. P. D. Taylor, L. Fahrig, K. Henein, G. Merriam, Connectivity Is a Vital Element of Landscape Structure. Oikos **68**, 571 (1993).

75. A. Gordon, et al., The use of dynamic landscape metapopulation models for forest management: a case study of the red-backed salamander. Canadian Journal of Forest Research **42**, 1091–1106 (2012).

76. N. C. R. Cadenhead, M. R. Kearney, D. Moore, S. Mcalpin, B. A. Wintle, Climate and Fire Scenario Uncertainty Dominate the Evaluation of Options for Conserving the Great Desert Skink. Conservation Letters (2015).

77. UN General Assembly, "Transforming Our World: The 2030 Agenda for Sustainable Development" (2015) (March 6, 2019).

**Supplementary Information for:**

**Assessing biophysical and socio-economic impacts of climate change on avian biodiversity**

**Authors:** Simon Kapitza, Pham Van Ha, Tom Kompas, Nick Golding, Natasha C. R. Cadenhead, Payal Bal and Brendan A. Wintle

**Corresponding author:** Simon Kapitza; simon.kapitza.research@gmail.com

**This PDF file includes:**

Supplementary Tables 1–5

Supplementary Figures 1–3

References for Supplementary Information

**Supplementary Table 1** | Climate, other biophysical, and socioeconomic predictors used as initial input to bias model, SDM, and land use model. Predictor choices were made based on literature. The initial predictor sets were reduced using correlation analysis.

| Short name | Long name | Chosen? | | | Source |
|---|---|---|---|---|---|
| | | Bias | SDM | Land use | |
| | **Climate predictors** | | | | (1) |
| *bio1* | Annual mean temperature | | x | x | |
| *bio2* | Mean diurnal range | | x | x | |
| *bio3* | Isothermality | | x | x | |
| *bio4* | Temperature seasonality | | x | x | |
| *bio5* | Maximum temperature of warmest month | | x | x | |
| *bio6* | Minimum temperature of coldest month | | x | x | |
| *bio7* | Temperature annual range | | x | x | |
| *bio8* | Mean temperature of wettest quarter | | x | x | |
| *bio9* | Mean temperature of driest quarter | | x | x | |
| *bio10* | Mean temperature of warmest quarter | | x | x | |
| *bio11* | Mean temperature of coldest quarter | | x | x | |
| *bio12* | Annual precipitation | | x | x | |
| *bio13* | Precipitation of wettest week | | x | x | |
| *bio14* | Precipitation of driest week | | x | x | |
| *bio15* | Precipitation of driest month | | x | x | |
| *bio16* | Precipitation of wettest quarter | | x | x | |
| *bio17* | Precipitation of driest quarter | | x | x | |
| *bio18* | Precipitation of warmest quarter | | x | x | |
| *bio19* | Precipitation of coldest quarter | | x | x | |
| | **Other biophysical predictors** | | | | |
| *roughness* | Roughness | x | | x | (2) |
| *slope* | Slope | | x | x | (2) |
| *srtm* | Elevation | | x | x | (2) |
| *diri* | Distance to Rivers | | x | x | (3) |
| *dila* | Distance to Lakes | | x | x | (3) |
| *dico* | Distance to Coast | | | x | (3) |
| *nitro* | Soil Nitrogen Content | | | x | (4) |
| *sawc* | Soil Available Water Content | | | x | (4) |
| *carb* | Soil Carbon Density | | | x | (4) |
| *bulk* | Soil Bulk Density | | | x | (4) |
| | **Socio-economic predictors** | | | | |
| *pa* | Protected Area | x | | x | (5) |
| *diro* | Distance to Roads | x | | x | (6) |
| *dibu* | Distance to Built-up Areas | x | | x | (7) |
| *popdi* | Population density | x | | | (8) |
| *landuse* | Land use – Urban | | x | x | (9) |
| | Land use – Cropland | | x | x | |
| | Land use – Herbaceous vegetation | | x | x | |
| | Land use – Shrubs | | x | x | |
| | Land use – Open Forest | | x | x | |
| | Land use – Closed Forest | | x | x | |
| | Land use - Herbaceous wetlands, moss and lichen | | x | | |
| | Land use - Bare soil and sparse vegetation | | x | | |
| *bioregions* | Bioregions in Australia | | x | | (10) |

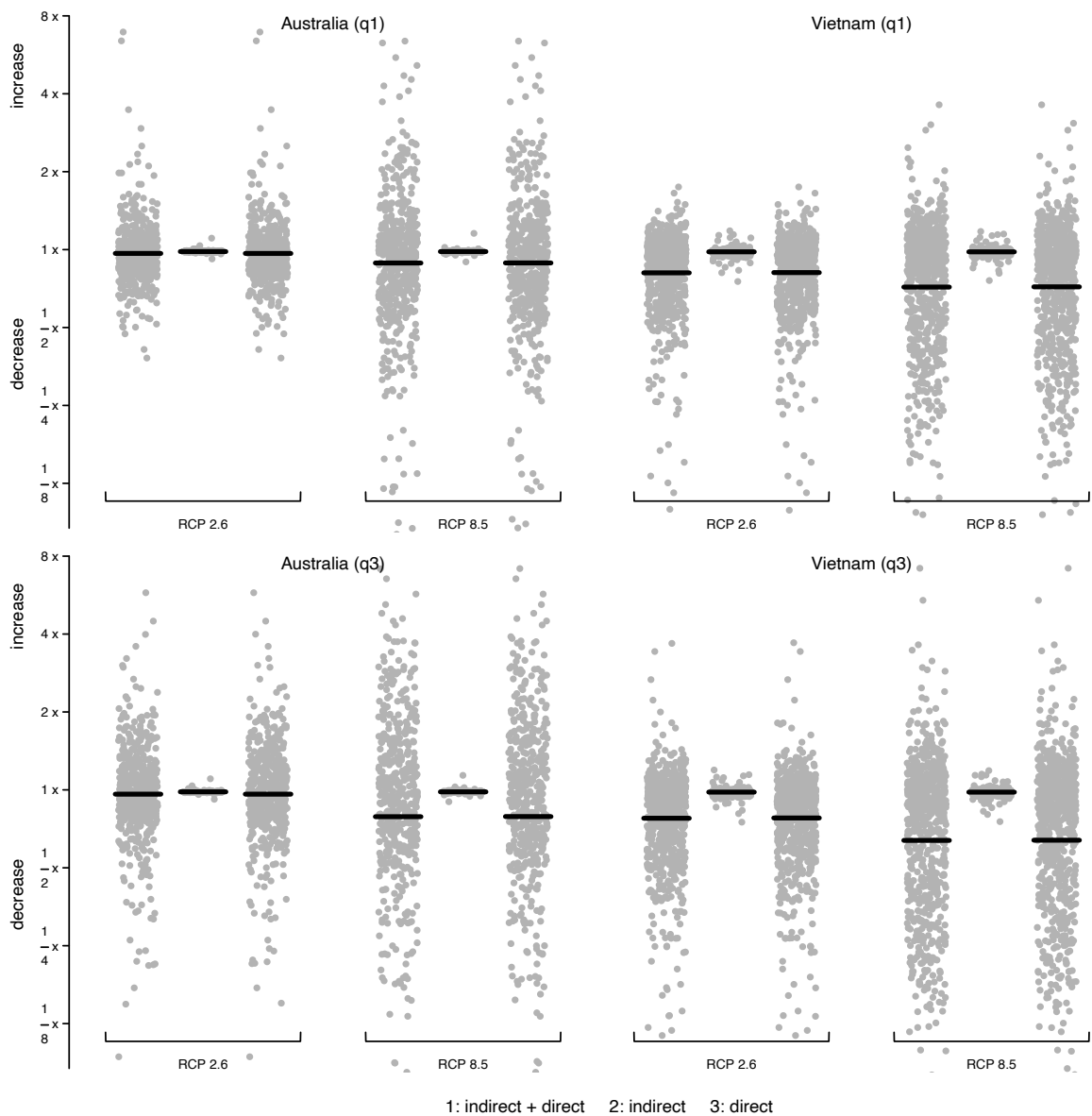

**Supplementary Figure 1** | Predicted change in mean potential habitat in Australia (**a** and **c**) and Vietnam (**b** and **d**) under the first quartile of GCMs (**a** and **c**) and the third quartile of GCMs (**c** and **d**). Main results are predictions under the cell-level medians. Plots are constrained on the y-axis to < 8x and > 1/16 x for visual clarity.

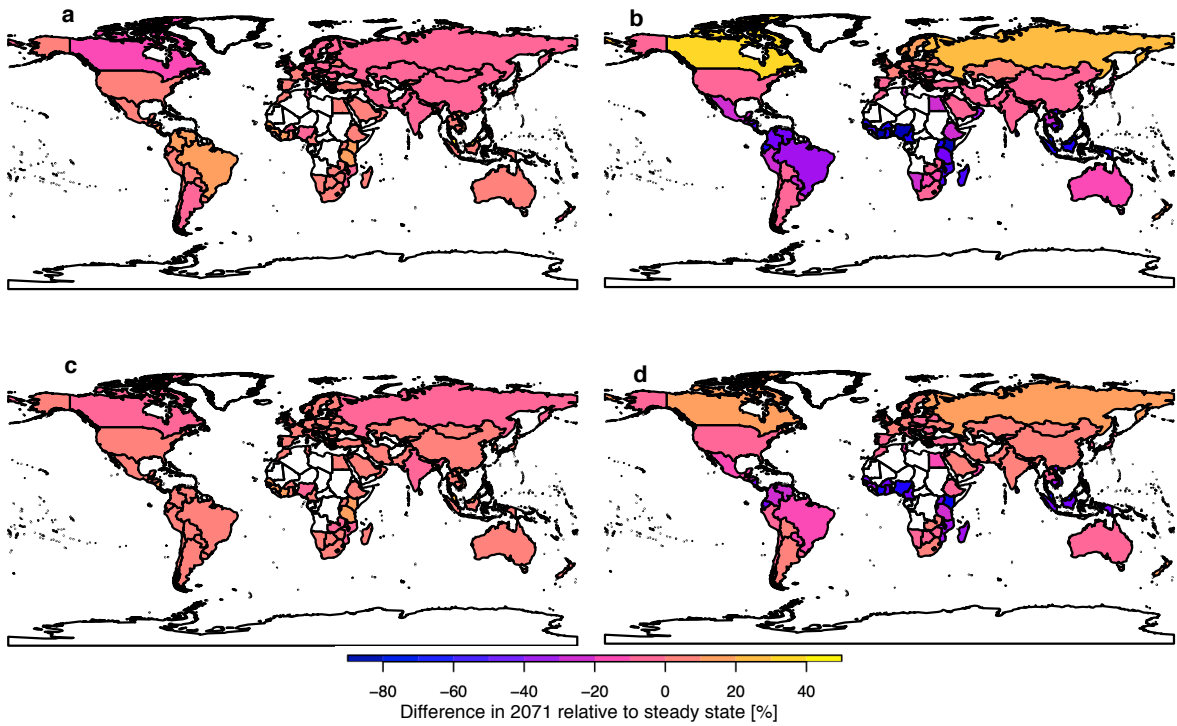

**Supplementary Figure 2 |** Relative changes in total wheat sector outputs (**a**, **b**) and land endowments (**c**, **d**) of GTAP 9 countries under RCP 2.6 (**a**, **c**) and RCP 8.5 (**b**, **d**). The % changes are relative to the economy after a forward propagation of the current economy without any scenario assumptions.

**Supplementary Table 2 | List of Global Circulation Models (GCM).** Downscaled outputs from these models were used to estimate cell-level medians and first and third quartiles of cell-level predictions of 19 biolcim variables. Data are available through WorldClim (11)

| GCM | Source |
|---|---|
| BCC-CSM1-1 | Beijing Climate Center, China Meteorological Administration |
| CCSM4 | University of Miami - RSMAS |
| CNRM-CM5 | Centre National de Recherches Météorologiques / Centre Européen de Recherche et Formation Avancée en Calcul Scientifique |
| GFDL-CM3 | NOAA Geophysical Fluid Dynamics Laboratory |
| GFDL-ESM2G | NOAA Geophysical Fluid Dynamics Laboratory |
| GISS-E2-R | NASA Goddard Institute for Space Studies |
| HadGEM2-AO | Met Office Hadley Centre (additional HadGEM2-ES realizations contributed by Instituto Nacional de Pesquisas Espaciais) |
| HadGEM2-ES | Met Office Hadley Centre (additional HadGEM2-ES realizations contributed by Instituto Nacional de Pesquisas Espaciais) |
| IPSL-CM5A-LR | Institut Pierre-Simon Laplace |
| MIROC-ESM-CHEM | Japan Agency for Marine-Earth Science and Technology, Atmosphere and Ocean Research Institute (The University of Tokyo), and National Institute for Environmental Studies |
| MIROC-ESM | Japan Agency for Marine-Earth Science and Technology, Atmosphere and Ocean Research Institute (The University of Tokyo), and National Institute for Environmental Studies |
| MIROC5 | Atmosphere and Ocean Research Institute (The University of Tokyo), National Institute for Environmental Studies, and Japan Agency for Marine-Earth Science and Technology |
| MPI-ESM-LR | Max-Planck-Institut für Meteorologie |
| MRI-CGCM3 | Meteorological Research Institute |
| NorESM1-M | Norwegian Climate Centre |

**Supplementary Table 3 | Relative contributions of commodity sectors to the total area harvested in 2016** (12). These values inform the weight of predicted land endowment changes when estimating total changes to crop land area.

| Sector | Full name | Australia | Vietnam |
|---|---|---|---|
| c_b | sugar cane, sugar beet | 0.020 | 0.018 |
| gro | cereal grains | 0.250 | 0.081 |
| ocr | crops nec | 0.094 | 0.129 |
| osd | oil seeds | 0.108 | 0.034 |
| pdr | paddy rice | 0.001 | 0.544 |
| pfb | plant-based fibres | 0.012 | 0.001 |
| v_f | vegetables, fruit, nuts | 0.019 | 0.193 |
| wht | wheat | 0.496 | - |

a

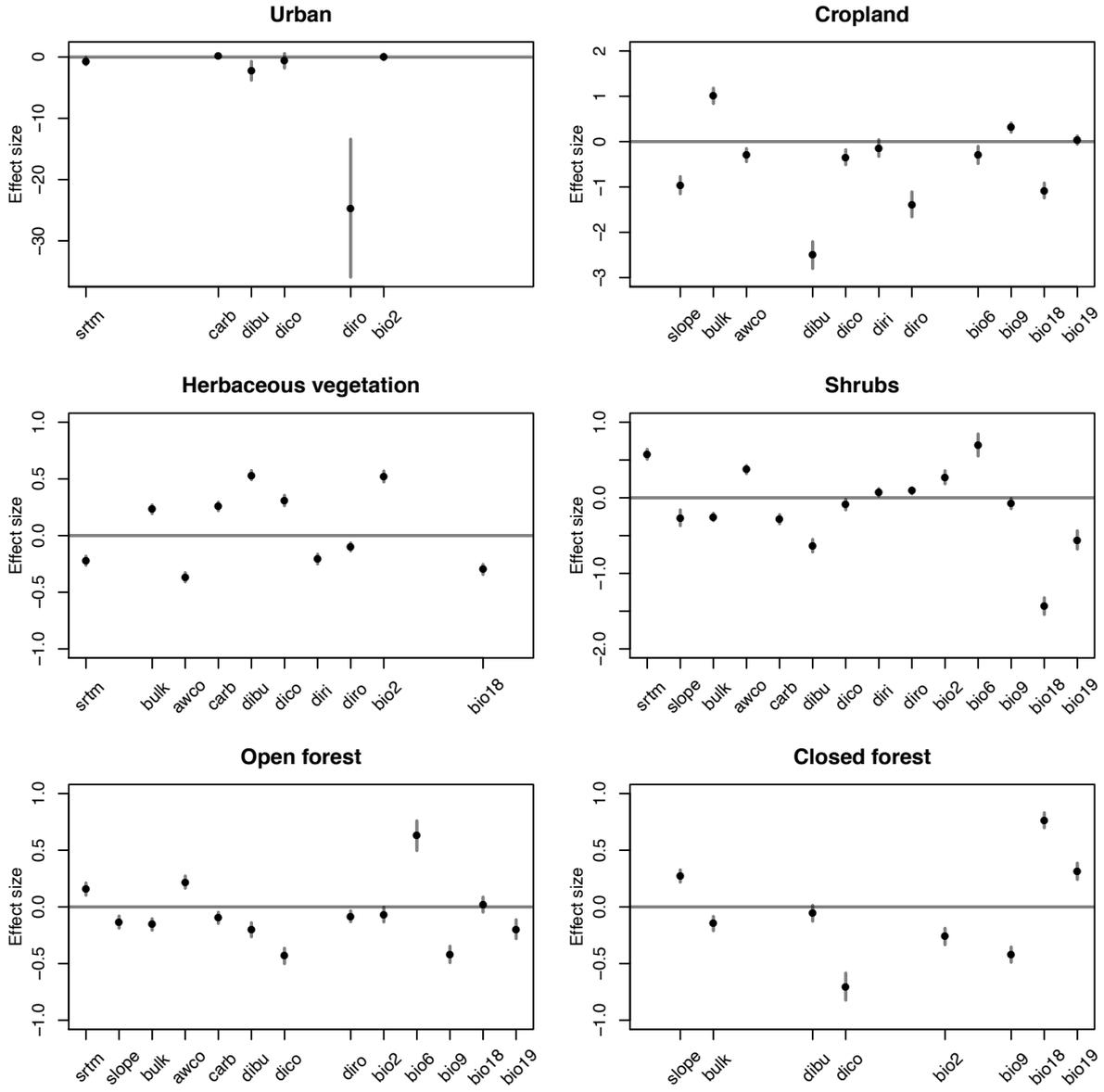

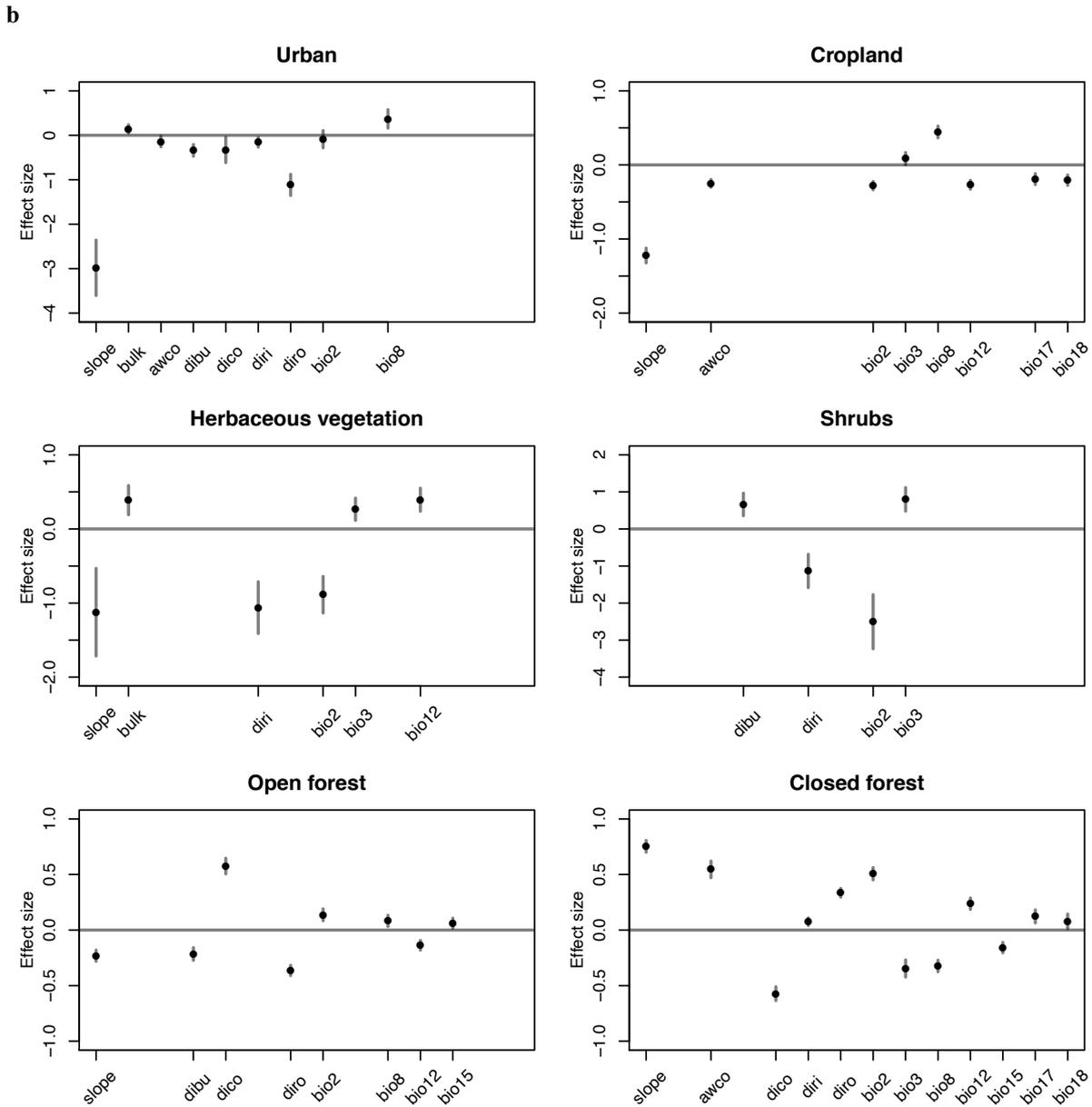

**Supplementary Figure 3 | Effect sizes of predictors in land use suitability models. a,** effect sizes of predictors in Australia and **b,** effect sizes of predictors in Vietnam. Predictors were standardised and we used cross-validated Lasso penalization for predictor selection. Error bars are indicated in grey. The sample size for model building was n = 20,000 in both countries.

**Supplementary Table 4 | Transition matrix of land use model.** 1 indicate possible transitions from the class of the row to the class of the column of the cell. 0 indicate when transitions are not possible.

| Class | Urban | Cropland | Herbaceous Ground Vegetation | Shrubs | Open Forest | Closed Forest |
|---|---|---|---|---|---|---|
| **Urban** | 1 | 0 | 0 | 0 | 0 | 0 |
| **Cropland** | 1 | 1 | 1 | 1 | 1 | 1 |
| **Herbaceous Ground Vegetation** | 1 | 1 | 1 | 1 | 1 | 1 |
| **Shrubs** | 1 | 1 | 1 | 1 | 1 | 1 |
| **Open Forest** | 1 | 1 | 1 | 1 | 1 | 1 |
| **Closed Forest** | 1 | 1 | 1 | 1 | 1 | 1 |

## References

1. R. J. Hijmans, S. E. Cameron, J. L. Parra, P. G. Jones, A. Jarvis, Very high resolution interpolated climate surfaces for global land areas. *International Journal of Climatology* **25**, 1965–1978 (2005).

2. NASA Land Processes Distributed Active Archive Center, "Shuttle Radar Topography Mission (SRTM) 1 Arc-Second Global" in (2000).

3. P. Wessel, W. H. F. Smith, A global, self-consistent, hierarchical, high-resolution shoreline database. *Journal of Geophysical Research: Solid Earth* **101**, 8741–8743 (1996).

4. Global Soil Data Task Group, Global Gridded Surfaces of Selected Soil Characteristics (IGBP-DIS) (2000) https:/doi.org/10.3334/ornldaac/569.

5. IUCN and UNEP-WCMC, The World Database on Protected Areas (WDPA) [online] [June 2018]. Cambridge, UK: UNEP-WCMC.

6. Center for International Earth Science Information Network - CIESIN - Columbia, Global Roads Open Access Data Set, Version 1 (gROADSv1) (2013).

7. FAO (Food and Agricultural Organisation, Built-up Areas of the World (Vmap0). **First edit** (1997).

8. Center for International Earth Science Information Network (CIESIN) Columbia University, International Food Policy Research Institute (IFPRI), The World Bank, Centro Internacional de Agricultura Tropical (CIAT), Global Rural-Urban Mapping Project, Version 1 (GRUMPv1): Population Density Grid. (2011).

9. European Union, "Copernicus Land Monitoring Service" (2019).

10. Department of Sustainability, Environment, Water, Population and Communities, Interim Biogeographic Regionalisation for Australia (IBRA), Version 7 (Regions). Bioregional Assessment Source Dataset. (September 28, 2018).

11. R. Hijmans, S. Cameron, J. Parra, P. Jones, A. Jarvis, WORLDCLIM - a set of global climate layers (climate grids), version 1.4.

12. Food and Agriculture Organization of the United Nations (FAO), *FAOSTAT Statistics Database* (2017) (September 18, 2018).